\documentclass[12pt]{article}

\usepackage{epsfig}
\input epsf

\title{Heavy quark and quarkonium production\\ at CERN LEP2:
$k_T$-factorization versus data}

\author{A.V.~Lipatov, N.P.~Zotov}

\begin{document}

\maketitle

\begin{center}

{\it D.V.~Skobeltsyn Institute of Nuclear Physics,\\ 
M.V. Lomonosov Moscow State University,
\\119992 Moscow, Russia\/}\\[3mm]

\end{center}

\vspace{1cm}

\begin{center}

{\bf Abstract }

\end{center}

We present calculations of heavy quark and quarkonium
production at CERN LEP2 in the $k_T$-factorization QCD approach. Both direct and 
resolved photon contribution are taken into account. 
The conservative error analisys is performed. The unintegrated gluon distribution in 
the photon is taken from the full CCFM evolution equation. The traditional 
color-singlet mechanism to describe non-perturbative transition of 
$Q\bar Q$-pair into a final quarkonium is used. Our analisys covers 
polarization properties of heavy quarkonia at moderate and large transverse momenta. 
We find that the total and differential open charm production cross sections 
are consistent with the recent experimental data taken by the L3, OPAL and ALEPH
collaborations. At the same time the DELPHI data for the 
inclusive $J/\psi$ production exceed our predictions but experimental 
uncertainties are too large to claim a significant inconsistency. 
The bottom production in photon-photon collisions at CERN LEP2 is hard to explain 
within the $k_T$-factorization formalism.

\vspace{1cm}

\section{Introduction} \indent 

Heavy quark and quarkonium production in photon-photon collisions at high energies 
is a subject of the intensive studies from both theoretical and experimental point~[1--6].
From the theoretical point, heavy quark in $\gamma \gamma$ collisions can be produced 
via direct and resolved production mechanisms. In direct events, the two photons couple directly to 
a heavy quark pair. In resolved events, one photon ("single-resolved") or both 
photon ("double-resolved") fluctuate into a hadronic state and a gluon or a quark
of this hadronic fluctuation takes part in the hard interaction. At LEP2 conditions 
the contribution of the double-resolved events ($gg \to Q\bar Q$) is small~[7], and
charm and bottom quarks are produced mainly via direct ($\gamma \gamma \to Q\bar Q$)
and single-resolved ($\gamma g \to Q\bar Q$) processes. The direct contribution
is not dependent on the quark and gluon content in the photon, whereas the 
single-resolved processes strongly depend on the photon's gluon density. Therefore 
detailed knowledge of gluon distributions in the photon is necessary for the 
theoretical description of such processes at modern (LEP2) and future (TESLA) colliders.

Usually quark and gluon densities are described by the 
Dokshitzer-Gribov-Lipatov-Altarelli-Parizi (DGLAP) evolution equation~[8] 
where large logarithmic terms proportional to $\ln(\mu^2)$ are taken into 
account. The cross sections can be rewritten in terms of process-dependent
hard matrix elements convoluted with quark or gluon density functions.
In this way the dominant contributions come from diagrams where parton emissions 
in initial state are strongly ordered in virtuality. This is called collinear 
factorization, as the strong ordering means that the virtuality of the parton 
entering the hard scattering matrix elements can be neglected compared to the 
large scale $\mu^2$. However, at the high energies this hard scale is large 
compare to the $\Lambda_{\rm QCD}$ parameter but on the other hand $\mu^2$ is 
much less than the total energy $\sqrt s$ (around 200 GeV for 
the LEP2 collider). Therefore in such case it was expected that the DGLAP 
evolution, which is only valid at large $\mu^2$, should break down. The 
situation is classified as "semihard"~[9--12].

It is believed that at assymptotically large energies (or small $x \sim \mu^2/s$)
the theoretically correct description is given by the Balitsky-Fadin-Kuraev-Lipatov
(BFKL) evolution equation~[13] because here large terms proportional to $\ln(1/x)$ are 
taken into account. Just as for DGLAP, in this way it is possible to factorize
an observable into a convolution of process-dependent hard matrix elements
with universal gluon distributions. But as the virtualities (and transverse 
momenta) of the propagating gluons are no longer ordered, the matrix 
elements have to be taken off-shell and the convolution made also over 
transverse momentum ${\mathbf k}_T$ with the unintegrated ($k_T$-dependent) gluon 
distribution ${\cal F}(x,{\mathbf k}_T^2)$. The unintegrated gluon distribution
${\cal F}(x,{\mathbf k}_T^2)$ determines the probability to find a gluon carrying the 
longitudinal momentum fraction $x$ and the transverse momentum ${\mathbf k}_T$. 
This generalized factorization is called $k_T$-factorization~[10, 11]. 

The unintegrated gluon distribution is a subject of intensive studies~[14]. 
Various approaches to investigate this quantity have
been proposed. One such approach, valid for both small and large
$x$, have been developed by Ciafaloni, Catani, Fiorani and Marchesini,
and is known as the CCFM model~[15]. It introduces angular ordering of emissions
to correctly treat gluon coherence effects. In the limit of 
asymptotic energies, it almost equivalent to BFKL~[16--18], but also similar to
the DGLAP evolution for large $x$ and high $\mu^2$. The resulting unintegrated
gluon distribution depends on two scales, the additional scale ${\bar q}^2$
is a variable related to the maximum angle allowed
in the emission and plays the role of the evolution scale $\mu^2$ in the
collinear parton densities. We will use the following classification
scheme~[14]: ${\cal F}(x,{\mathbf k}_T^2)$ is used for pure BFKL-type unintegrated 
gluon distributions and ${\cal A}(x,{\mathbf k}_T^2,\mu^2)$ stands for any other 
type having two scale involved.

The CCFM evolution equation formulated for the 
proton has been solved numerically or analitically by different ways~[19--21].
As it was shown~[22--25], unintegrated gluon distribution in the photon ${\cal A}_{\gamma}(x,{\mathbf k}_T^2,\mu^2)$ 
can be constructed by the similar method as in the proton\footnote{See also~[26], where
we have used for unintegrated gluon density in a photon the prescription 
proposed by J.~Bl\"umlein~[27].}. But situation is 
a bit different due to the pointlike component which reflects the splitting of the
photon into a quark-antiquark pair. Also in the photon there are no sum rules equivalent
to those in the proton case that constrain the quark distributions.
However, this difference is not significant because CCFM equation 
contains only gluon splitting $g \to gg$.
For the first time the unintegrated gluon density in the 
photon was obtained~[22] using a simplified 
solution of the CCFM equation in single loop approximation, when small-$x$ effects 
can be neglected. It means that the CCFM evolution is reduced to the DGLAP one with 
the difference that the single loop evolution takes the gluon transverse momentum 
$k_T$ into account. Another simplified solution of the CCFM equation for a photon 
was proposed~[23] using the Kimber-Martin-Ryskin (KMR) prescription~[20]. In this way 
the $\mu^2$-dependence in the unintegrated gluon distribution enters only in 
last step of the evolution, and one-scale evolution equations can be used up to 
the last step. Both these methods give the similar results~[23]. 
The phenomenological unintegrated gluon density, based on the 
Golec-Biernat and Wusthoff (GBW) saturation model~[24] (extended to the 
large-$x$ region), was proposed also~[23]. Very recently the unintegrated gluon distribution 
in the photon ${\cal A}_{\gamma}(x,{\mathbf k}_T^2,\mu^2)$ was obtained~[25] using the full 
CCFM evolution equation for the first time. It was shown that the full CCFM 
evolved effective (integrated over ${\mathbf k}_T^2$) distribution is much 
higher than the usual DGLAP-based gluon density at $x < 10^{-1}$ region.

In the previous studies~[23, 25] the unintegrated gluon distributions in a photon 
(obtained from the single loop as well as full CCFM evolution equation) were applied
to the calculation of the open charm and bottom production at LEP2.
It was found that total cross section of the charm production is
consistent with experimental data. In contrast, the theoretical predictions 
of the bottom total cross section underestimate data by factor 2 or 3.
But we note that all these calculations reveal to the total cross sections of 
the open charm and bottom production only. In this paper we will 
study heavy flavored production at LEP2 more detail using the 
full CCFM-evolved unintegrated gluon density~[25].
We will investigate the total and differential heavy quark cross section 
(namely pseudo-rapidity and transverse momenta distributions of the 
$D^*$-meson) and compare our theoretical results with the recent experimental 
data taken by the L3, OPAL and ALEPH collaborations at LEP2~[1--5].

Also we will study here the very intriguing problem connected with the
inclusive $J/\psi$ meson production at high energies. It is traces
back to the early 1990s, when the CDF data on the $J/\psi$ and $\Upsilon$
hadroproduction cross section revealed a more than order of magnitude discrepancy
with theoretical expectations. This fact has resulted to intensive theoretical 
investigations of such processes. In the so-called non-relativistic QCD (NRQCD)~[28] 
there are additional (octet) transition mechanism from $c\bar c$ pair to the 
$J/\psi$ mesons, where $c\bar c$ pair is produced in the color octet (CO) state and 
transforms into final color singlet state (CS) by help soft gluon radiation.
The CO model describes well the heavy quarkonium production at Tevatron~[29], although
there are also some indications that it does not work well.
For example, contributions from the octet mechanism to the $J/\psi$ photo- and leptoproduction
processes at HERA are not well reproduce~[30] experimental data. Also NRQCD is not predict 
$J/\psi$ polarization properties at HERA and Tevatron~[30, 31].
At the same time usual CS model supplemented with $k_T$-factorization formalism gives fully
correct description of the inclusive $J/\psi$ production at HERA~[32] and Tevatron~[33] including 
spin alignement of the quarkonium.
The CO contributions within the $k_T$-factorization approach also will not 
contradict experimental data if parameters of the non-perturbative matrix elements 
will be reduced~[33, 34]. But this fact changes hierarchy of these matrix elements 
which are obtained within the NRQCD.

Recently DELPHI collaboration has presented experimental data~[35]
on the inclusive $J/\psi$ production in $\gamma \gamma$ collisions at LEP2,
which wait to be confronted with different theoretical predictions.
The theoretical calculations~[36] within the NRQCD formalism agree well with the
DELPHI data. In contrast, the collinear DGLAP-based leading order perturbative 
QCD calculations in the CS model significantly (by order of magnitude) underestimate~[36] the data. 
The aim of this paper, in particular, is to investigate whether the inclusive $J/\psi$
production at LEP2 can be explained in the traditional CS model by using 
$k_T$-factorization and CCFM-based unintegrated gluon density in a photon.

The outline of this paper is following. In Section 2 we present the basic
formalism of the $k_T$-factorization approach with a brief review of calculation steps. 
In Section 3 we present the numerical results of
our calculations. Finally, in Section 4, we give some conclusion.
The compact analytic expressions for the off-shell matrix elements of the 
all subprocesses under consideration are given in Appendix. These formulas 
may be useful for the subsequent applications. 

\section{Cross sections for heavy flavour production} \indent 

In $\gamma \gamma$ collisions heavy quark and quarkonium can be produced by one of the 
three mechanisms: a direct production, a single-resolved and a double-resolved
production processes. The direct contribution is governed by simple QED
amplitudes which is independent on the gluon density in the photon.
The double-resolved process gives a much smaller contribution than the direct and
single-resolved processes~[7] and will not taken into account in this analysis.

Let $p_{\gamma}^{(1)}$ and $p_{\gamma}^{(2)}$ be the momenta of the incoming
photons and $p_1$ and $p_2$ the momenta of the produced quarks.
The single-resolved contribution to the $\gamma \gamma \to Q\bar Q$ process 
in the $k_T$-factorization approach has the following form:
$$
  d\sigma_{\rm 1-res}(\gamma \gamma \to Q\bar Q) = \int {dx\over x} {\cal A}_{\gamma}(x,{\mathbf k}_T^2,\mu^2) 
  d{\mathbf k}_T^2 {d\phi \over 2\pi} d \hat \sigma(\gamma g^* \to Q\bar Q), \eqno (1)
$$

\noindent
where $\hat \sigma(\gamma g^* \to Q\bar Q)$ is the heavy quark production
cross section via off-shell gluon having fraction $x$ of a photon longitudinal 
momentum, non-zero transverse momentum 
${\mathbf k}_T$ (${\mathbf k}_T^2 = - k_T^2 \neq 0$) and azimuthal angle $\phi$. 
The expression (1) can be easily rewritten as
$$
  { d\sigma_{\rm 1-res} (\gamma \gamma \to Q\bar Q) \over dy_2 d{\mathbf p}_{2T}^2 } = \int {1\over 16\pi (x s)^2 (1 - \alpha_2)} {\cal A}_{\gamma}(x,{\mathbf k}_T^2,\mu^2) 
  |\bar {\cal M}|^2(\gamma g^* \to Q\bar Q) d{\mathbf k}_T^2 {d\phi \over 2\pi} {d\phi_2 \over 2\pi}, \eqno (2)
$$

\noindent
where $|\bar {\cal M}|^2(\gamma g^* \to Q\bar Q)$ is the off-shell matrix element, 
$s = (p_{\gamma}^{(1)} + p_{\gamma}^{(2)})^2$ is the total c.m. frame energy, $y_2$ and $\phi_2$ are the rapidity
and azimuthal angle of the produced heavy quark having mass $m_Q$, 
$\alpha_2 = m_{2T}\exp(y_2)/\sqrt s$ and $m_{2T}^2 = m_Q^2 + {\mathbf p}_{2T}^2$.
To calculate the single-resolved contribution to the inclusive $J/\psi$ production 
the same formula (2) can be used where off-shell maxtrix element 
$|\bar {\cal M}|^2(\gamma g^* \to Q\bar Q)$ should be replaced by one which corresponds
to the $\gamma g^* \to J/\psi + g$ production process.

The available experimental data~[1--5, 35] refer to heavy quark or quarkonium 
production in the $e^+ e^-$ collisions also. In order to obtain
these total cross sections, the $\gamma \gamma$ cross sections need to
be weighted with the photon flux in the electron:
$$
  d\sigma(e^+ e^- \to e^+ e^- Q\bar Q + X) = \int f_{\gamma/e}(x_1)dx_1 \int f_{\gamma/e}(x_2)dx_2\,d\sigma(\gamma \gamma \to Q\bar Q), \eqno (3)
$$

\noindent
where we use the Weizacker-Williams approximation for the bremsstrahlung photon
distribution from an electron:
$$
  f_{\gamma/e}(x) = {\alpha \over 2\pi}\left({1 + (1 - x)^2\over x}\ln{Q^2_{\rm max}\over Q^2_{\rm min}} + 
  2m_e^2 x\left({1\over Q^2_{\rm max}} - {1\over Q^2_{\rm min}} \right)\right), \eqno (4)
$$

\noindent
with $Q^2_{\rm min} = m_e^2x^2/(1 - x)^2$ and $Q^2_{\rm max} = Q^2_{\rm min} + (E\theta)^2(1 - x)$.
Here $x = E_{\gamma}/E_e$, $E = E_e = {\sqrt s}/2$, $s$ is the total energy in the 
$e^+ e^-$ collision and $\theta \sim 30$ mrad is the angular cut that ensures
the photon is quasi-real.

The multidimensional integration in (2) and (3) has been performed
by means of the Monte Carlo technique, using the routine VEGAS~[37].
The full C$++$ code is available from the authors on request\footnote{lipatov@theory.sinp.msu}.
For reader's convenience, we collect the analytic expressions for the 
off-shell matrix elements $|\bar {\cal M}|^2(\gamma g^* \to Q\bar Q)$ and 
$|\bar {\cal M}|^2(\gamma g^* \to J/\psi g)$ in Appendix, including, in particular, 
the relevant formulas for helicity zero $J/\psi$ production state.

\section{Numerical results} \indent 

First of all, we can perform integration of the unintegrated gluon 
distribution ${\cal A}_{\gamma}(x,k_T^2,\mu^2)$~[25] over gluon 
transverse momenta ${\mathbf k}_T^2$ to obtain the effective gluon density in the photon:
$$
  xg_{\gamma}(x,\mu^2) \sim \int\limits_0^{\mu^2} {\cal A}_{\gamma}(x,{\mathbf k}_T^2,\mu^2) d{\mathbf k}_T^2. \eqno(5)
$$

\noindent
The effective density $xg_{\gamma}(x,\mu^2)$ can be compared with
the experimental data~[38, 39] taken by H1 collaboration at HERA. As seen in Figures 1 and 2,
this gluon distribution agrees well to the existing data extracted from the
hard dijet (mean $p_T^2 = 38\,{\rm GeV}^2$~[38] and $p_T^2 = 74\,{\rm GeV}^2$~[39]).
In contrast, KMR construction of unintegrated gluon density in the photon
tends to underestimate the HERA data at $x < 10^{-1}$~[23]. 

Being sure that the 
full CCFM-evolved unintegrated gluon density ${\cal A}_{\gamma}(x,k_T^2,\mu^2)$ 
reproduces well the experimental 
data for the $xg_{\gamma}(x,\mu^2)$, we now are in a position to present our numerical results.
We describe first our theoretical input and the kinematical 
conditions. The cross sections for heavy quark and quarkonium production depend on
the heavy quark mass and the energy scale $\mu^2$. 
Since there are no free quarks due to confinement effect, their masses 
cannot be measured directly and should be defined from hadron properties. 
In our analysis we have examined the following choice: $m_c = 1.4 \pm 0.1$ GeV for 
charm and $m_b = 4.75 \pm 0.25$ GeV for bottom quark masses. Such variation of the quark masses 
gives the largest uncertainties in comparison with scale variation\footnote{It was 
shown~[25] that Monte Carlo generator CASCADE~[19, 40] predicts the very similar results for charm total cross section with
both scales $\mu^2 = m_Q^2 + {\mathbf p}_{T}^2$ and $\mu^2 = 4 m_Q^2$.}
and therefore can be used as an estimate of the total theoretical uncertainties. 
Then, we will apply standard expression $\mu^2 = m_Q^2 + {\mathbf p}_{T}^2$ for both 
renormalization and factorization scales. 
Here ${\mathbf p}_{T}$ is the transverse momentum of the heavy quark in the center-of-mass frame.
We use LO formula for the strong coupling constant 
$\alpha_s(\mu^2)$ with $n_f = 4$ active quark flavours and
$\Lambda_{\rm QCD} = 200$ MeV, such that $\alpha_s(M_Z^2) = 0.1232$.
But $\Lambda_{\rm QCD} = 340$ MeV~[7] was tested also.

Figure 3 confronts the total cross section $\sigma(\gamma \gamma \to c\bar c + X)$
calculated as a function of the photon-photon total energy 
$W_{\gamma \gamma}$ with experimental data~[2] taken by the L3 collaboration in the 
interval $5 < W_{\gamma \gamma} < 70$ GeV. Solid line represent the 
calculations with the charm mass $m_c = 1.4$ GeV, whereas upper and lower 
dashed lines correspond to the $m_c = 1.3$ GeV and $m_c = 1.5$ GeV respectively.
It is clear that $k_T$-factorization reproduces well both the 
energy dependence and the normalization. One can see that sensitivity of
the results to the variations of the charm mass is rather large: shifting the
mass down to $m_c = 1.3$ GeV changes the estimated cross section 
by $15 - 20$\% at $W_{\gamma \gamma} \sim 60$ GeV. But in general all three
curves lie within the experimental uncertainties.

Experimental data for the total cross section $\sigma(e^+ e^- \to e^+ e^- c\bar c + X)$
come from the three LEP collaborations L3~[1, 3], OPAL~[4] and ALEPH~[5]. In Figure 4 we
show our predictions in comparison with data. All curves here are the same as in Figure 3.
One can see that our calculations describe the experimental data well again. 
The variation of the quark masses $1.3 < m_c < 1.5$ GeV gives
the theoretical uncertainties approximately $15$\% in absolute normalization.

The available experimental data were obtained for the $D^{*}$ meson 
production also. Two differential cross section are determined: the first
one as a function of the transverse $D^*$ momentum $p_T$, and the second as
a function of pseudo-rapidity $|\eta|$. In our calculation we convert 
charmed quark into $D^*$ meson using the Peterson fragmentation function~[41]. 
The default set for the fragmentation parameter $\epsilon_c$ and the 
fraction $f(c \to D^*)$ is $\epsilon_c = 0.06$ and $f(c \to D^*) = 0.26$. 
Other values for the parameter $\epsilon_c$ in the NLO perturbative QCD 
calculations are often used also, namely $\epsilon_c = 0.116$~[42] in
the massless scheme and $\epsilon_c = 0.031$~[43] in the massive one. 
In the case of the massless calculation, this parameter was determined via a NLO fit
to LEP1 data on $D^*$ production in $e^+ e^-$ annihilation measured by OPAL collaboration~[44].
To investigate the sensitivity of the our numerical results to $\epsilon_c$ parameter
we have repeat our calculations using $\epsilon_c = 0.031$. 

The recent L3 data~[3] refer to the
kinematic region defined by $1 < p_T < 12$ GeV and $|\eta| < 1.4$
with averaged total $e^+ e^-$ energy $\sqrt s = 193$ GeV 
($183 < \sqrt s < 209$ GeV). The OPAL data~[4] were obtained
in the region $2 < p_T < 12$ GeV, $|n| < 1.5$ and averaged
over $183 < \sqrt s < 189$ GeV. The more recent ALEPH data~[5] refer to the
same kinematic region but averaged over $183 < \sqrt s < 209$ GeV.
We compare these three data sets with our calculation at $\sqrt s = 193$ GeV. 
The different values of $\sqrt s$ are not expected to change
the cross section more than the corresponding experimental errors.
We have checked directly that shifting $\sqrt s$ from 189 to 193 GeV increase
the calculated cross sections by about one percent only.

The transverse momenta distributions of the $D^*$ meson 
for different pseudo-rapidity region in comparison to experimental 
data shown in Figure 5 and 6. 
The solid and both dashed curves here are the same as in Figure 3 (calculated 
with the default value $\epsilon_c = 0.06$), dash-dotted curves represent results 
obtained using $\epsilon_c = 0.031$ and $m_c = 1.4$ GeV.
The overall agreement between the our predictions and the
data is good although the ALEPH data points in medium and large $p_T$ range
lie slightly above the theoretical curves. 
Shifting the $\epsilon_c$ default value down to $\epsilon_c = 0.031$
results to a bit broadening of the $p_T$ spectra, which is
insufficient to describe the data.
The effects come from changing of the charm mass present only at low $p_T$: the predicted
cross section $d\sigma/dp_T$ with $m_c = 1.3$ GeV is $10 - 15$\% above 
the one calculated with $m_c = 1.4$ GeV at $p_T \sim 1$ GeV, whereas 
solid and both dashed theoretical curves practically coincide at medium and large $p_T$.
The similar effect was found in the NLO perturbative calculation~[45], where
the difference between the massive and massless approach arises at low
$p_T$ only.

The $D^*$ pseudo-rapidity distributions compared with the 
experimental data in different $p_T$ range are shown in Figure 7 and 8.
All curves here are the same as in Figure 5. Our calculations 
agree well with measured differential cross sections but
slightly underestimate the OPAL data. However,
setting $\epsilon_c = 0.031$ increases the absolute normalization of the 
pseudo-rapidity distribution by approximately $10$\%, and agreement with OPAL
data becomes better. Again, one can see that the
significant mass dependence has place at low transverse momenta only:
the difference between the theoretical curves calculated at 
$2 < p_T < 12$ GeV and plotted in Figure 8 is much smaller than difference 
between the results presented in Figure 7 obtained at $1 < p_T < 12$ GeV.

We conclude from Figs.~3 --- 8, that our calculations agree well with charm data at LEP2.
In contrast with charm case, the open bottom production in $\gamma \gamma$ collisions 
is clear underestimated by $k_T$-factorization. Figure 9 shows our prediction for 
the open bottom cross section compared to the L3~[1] and OPAL~[4] experimental data.
Using low but still reasonable $b$-quark mass $m_b = 4.5$ GeV, we obtain 
$\sigma(e^+ e^- \to e^+ e^- b\bar b + X) = 2.94$ pb at $\sqrt s = 200$ GeV.
The very similar value $\sigma = 2.7$ pb was obtained~[23] 
within the GBW saturation model adopted for the photon. The Monte Carlo 
generator CASCADE predicts $\sigma = 4.9$ pb~[25] where normalization factor
$n = 1.7$ has been applied for the resolved contributions.
The calculation~[23] based on the KMR prescription for unintegrated gluon density
gives a lower cross section $\sigma = 1.9$ pb. At the same time the prediction
of the massive NLO perturbative QCD calculation~[43] 
is 3.88 pb for $m_b = 4.5$ GeV and 2.34 pb for $m_b = 5.2$ GeV
respectively. All these results are significantly (about factor 2 or 3) 
lower than experimental data. 

Such disagreement between theory and data for bottom production 
at LEP2 is surprising and needs an explanation.
It is known that the similar difference between theory and data was claimed for inclusive 
bottom hadroproduction at Tevatron. Recent analysis indicates that the overall
description of the these data can be improved~[46] by adopting the 
non-perturbative fragmentation function of the $b$-quark into 
$B$ meson: an appropriate treatment of the $b$-quark fragmentation 
properties considerably reduces the disagreement between measured bottom 
cross section and the corresponding NLO calculations. 
It would be interesting to find out whether the similar explanation is also true for the 
L3 and OPAL experimental data.

After we have studied open heavy quark production at LEP2, we will investigate 
production of the heavy quarkonum in $\gamma \gamma$ interactions. As it was already 
mentioned above, non-relativistic QCD gives a good description~[36] of the recent DELPHI 
data~[35] on inclusive $J/\psi$ production at LEP2. 
We will examine whether the DELPHI data can be expained
within the CS model using $k_T$-factorization approach and CCFM-based unintegrated gluon
density in a photon. Again, only direct and single-resolved contributions are taken into account.

Now we change the default set of parameters which were used in the case of 
open charm calculations. 
Since we neglect the relative momentum of the $c$-quarks (which 
form a $J/\psi$ meson) the charm mass should be taken $m_c = m_{\psi}/2$. Therefore 
as default choice in following we will use $m_c = 1.55$ GeV. On the
other hand there are many examples when smaller value $m_c = 1.4$ GeV 
in the calculation of $J/\psi$ production is used~[30, 32, 36]. In our analysis 
we will apply this value as extremal choice to investigate the theoretical
uncertainties of the calculations.

The DELPHI data~[35] refer to the kinematic region defined by $-2 < y_{\psi} < 2$ with
total $e^+ e^-$ energy $\sqrt s = 197$ GeV, where $y_{\psi}$ is the $J/\psi$ rapidity. 
However these data were obtained starting from very low
transverse momenta $p_{\psi T}^2 \sim 0.2\,{\rm GeV}^2$. We note that 
$k_T$-factorization as well as usual collinear factorization theorem does not work well
for such $p_{\psi T}$ values, and our calculations should be compared with 
experimental data at approximately $p_{\psi T} > 1$ GeV only.
In Figure 10 we confront our theoretical predictions with the measured differential 
cross section $d\sigma/dp_{\psi T}^2$. Solid line corresponds the default set of parameters and
lies below the data by a factor about 2 or 3. This dicrepancy is not catastrofic, because
some reasonable variations in $m_c$ and $\Lambda_{\rm QCD}$, namely 
$200 < \Lambda_{\rm QCD} < 340$ MeV and $1.4 < m_c < 1.55$ GeV, 
change the estimated cross section by a factor of 3 (dashed line in Figure 9).
Thus the visible disagreement is eliminated. However, we do not interpret this as a strong
indication of consistency between data and theory, but rather as a concequence 
of a wide uncertainty band. Better future experimental studies are crucial to
determine whether the results of our calculations contradict DELPHI data points.

The main difference between $k_T$-factorization and other approaches
connects with polarization properties of the final particles because the 
initial off-shell gluons do promptly manifest in the $J/\psi$ spin alignement~[32--34].
Only a very small fraction of $J/\psi$ mesons can be produced in the
helicity zero state (longitudinal polarization) by massless bosons. This
property is totally determined by the subprocess matrix element structure.
The degree of spin alignement can be measured experimentally since the
different polarization states of $J/\psi$ result in significantly different
angular distributions of the $J/\psi \to \mu^+ \mu^-$ decay leptons:
$$
  {d\Gamma(J/\psi \to \mu^+ \mu^-) \over d \cos\theta} \sim 1 + \alpha \cos^2\theta, \eqno (6)
$$

\noindent
where $\alpha = 1$ for transverse and $\alpha = - 1$ for 
longitudinal polarizations, respectively. Here $\theta$ is the angle between the 
lepton and $J/\psi$ directions, measured in $J/\psi$ meson rest frame. 
We calculate the $p_T^2$-dependence of the spin alignement parameter as
$$
  \alpha(p_{\psi T}^2) = {1 - 3 \xi(p_{\psi T}^2) \over 1 + \xi(p_{\psi T}^2)}, \eqno (7)
$$

\noindent
with $\xi(p_{\psi T}^2)$ being the fraction of longitudinally polarized
$J/\psi$ mesons. The results of our calculations are shown in Figure 11.
Solid line represents the $k_T$-factorization predictions and
dashed line corresponds to the collinear leading-order pQCD ones with GRV (LO) gluon
density~[47] in a photon.
One can see that fraction of longitudinally polarized mesons increases with $p_{\psi T}^2$
within $k_T$-factorization approach. This fact is in clear contrast with usual 
collinear parton model result.
The $k_T$-factorization calculations made for the inelastic $J/\psi$ production at HERA~[32, 34] 
and Tevatron~[33] posses the same behavior of the $\alpha$ parameter, whereas 
collinear parton model and NRQCD predict the strong transverse polarization at moderate and large 
$p_{\psi T}^2$ range. 
We point out that our predictions for the $J/\psi$ polarization are stable
with respect to variation of the model parameters, such as charmed quark mass and
factorization scale. In fact, there is no dependence on the strong coupling
constant which is cancels out.
At the same time the DELPHI fit~[35] gives $\alpha = 0.7 \pm 1.3$ for 
$p_{\psi T}^2 > 1\,{\rm GeV^2}$, which has, however, huge
experimental uncertainties. Since account of the octet contributions 
does not change predictions of the $k_T$-factorization approach for the spin 
alignement parameter $\alpha$~[33], the future extensive experimental study of  
such processes will be direct probe of the gluon virtuality.

\section{Conclusions} \indent 

We have investigated a heavy flavour production in photon-photon interactions
at high energies within the framework of $k_T$-factorization, using unintegrated
gluon distribution obtained from the full CCFM evolution equation for a photon.
We calculate total and differential cross sections of the open charm and bottom 
production including $D^*$ meson transverse momenta and pseudo-rapidity distributions. 
Also we have studied inclusive $J/\psi$ production at LEP2 using 
color-singlet model supplemented with $k_T$-factorization.

We take into account both the direct and single-resolved
contribution and investigate the sensitivity of the our results to the different 
parameters, such as heavy quark mass, charm fragmentation and 
$\Lambda_{\rm QCD}$ parameter. 
There are, of course, also some uncertainties due to the renormalization and
factorization scales. However, these effects would not to be large enough to change 
the conclusions presented here, and were not taken into account in our analisys.

The results of calculations with default parameter set agree well with open charm 
production data taken by
the L3, OPAL and ALEPH collaborations at LEP2. In contrast, bottom production
cross section is clearly underestimated by a factor about 3. A potential
explanation of this fact may be, perhaps, connected with the more 
accurate treatment of the $b$-quark 
fragmentation function. Our prediction for the inclusive $J/\psi$ production
slightly underestimate the DELPHI data. However, a strong inconsistensy
cannot be claimed, because of large experimental errors and theoretical 
uncertainties. Therefore more precise future experiments, espesially
polarized quarkonium production, are necessary to know whether 
our predictions contradict the experimental data.

In conclusion, we point out that at CERN LEP2 collider (as well as at HERA and Tevatron) the
difference between predictions of the collinear and $k_T$-factorization approaches is
clearly visible in polarized heavy quarkonium production. It comes directly
from initial gluon off-shellness.
The experimental study of such processes should be additional test of non-collinear
parton evolution.

\section{Acknowledgements} \indent 

We tnank Hannes Jung for possibility to use the CCFM code for unintegrated gluon 
distribution in a photon in our calculations.
The authors are very grateful also to Sergei Baranov for encouraging interest
and helpful discussions. One of us (A.L.) was supported by the 
INTAS grant YSF'2002 N 399 and by the "Dynasty" fundation.

\section{Appendix} \indent 

Here we present compact analytic expressions for the 
off-shell matrix elements which appear in (2). In the following,
$\hat s$, $\hat t$, $\hat u$ are usual Mandelstam variables for $2 \to 2$ process
and $e_Q$ is the fractional electric charge of heavy quark $Q$.

We start from photon-gluon
fusion $\gamma (q) g^* (k)\to Q (p_1) \bar Q (p_2)$ subprocess, where initial off-shell gluon 
has non-zero virtuality ${\mathbf k}_T^2$. The corresponding squared 
matrix element summed over final polarization states and averaged over initial ones read
$$
  |\bar {\cal M}|^2(\gamma g^* \to Q\bar Q) = - {\displaystyle (4\pi)^2 e_Q^2 \alpha \alpha_s \over 
  \displaystyle (\hat t - m^2)^2 (\hat u - m^2)^2 } F_{Q\bar Q}({\mathbf k}_T^2), \eqno (A.1)
$$

\noindent
where $m$ is the heavy quark mass, and 
$$
  F_{Q\bar Q}({\mathbf k}_T^2) = 6m^8 - (2{\mathbf k}_T^4 + 2(\hat t + \hat u){\mathbf k}_T^2 + 3{\hat t}^2 + 3{\hat u}^2 + 14 \hat t\hat u)m^4 + 
$$
$$
  (2(\hat t + \hat u){\mathbf k}_T^4 + 8\hat t \hat u{\mathbf k}_T^2 + {\hat t}^3 + {\hat u}^3 + 7 \hat t {\hat u}^2 + 7{\hat t}^2\hat u)m^2 - 
$$
$$
  \hat t\hat u(2{\mathbf k}_T^4 + 2(\hat t + \hat u){\mathbf k}_T^2 + {\hat t}^2 + {\hat u}^2). \eqno (A.2)
$$

\noindent
We note that matrix element of the direct contribution $\gamma \gamma \to Q\bar Q$ may be
easily obtained from (A.1) and (A.2) in the limit ${\mathbf k}_T^2 \to 0$, 
if we replace normalization factor $(4\pi)^2 e_Q^2 \alpha \alpha_s$ 
by the $(4\pi)^2 \alpha^2 e_Q^4 N_c$ (where $N_c$ is the number of colors) and average (A.2) 
over transverse momentum vector ${\mathbf k}_T$.

Now we are in a position to present our formulas for the 
$\gamma (q) g^* (k) \to J/\psi (p_\psi) g (p_g) $ subprocess.
In the color-singlet model the production of $J/\psi$ meson is considered 
as production of a quark-antiquark system in the color-singlet state with 
orbital momentum $L = 1$ and spin momentum $S = 1$. The squared
off-mass shell matrix element summed over final polarization states and averaged 
over initial ones can be written as
$$
  |\bar {\cal M}|^2(\gamma g^* \to J/\psi g) = - {64 e_Q^2 (4\pi)^3 \alpha \alpha_s^2 |\psi(0)|^2 \over 
  3 m_{\psi} (\hat t - m_\psi^2)^2 (\hat u - m_\psi^2 - {\mathbf k}_T^2)^2 (\hat t + \hat u + {\mathbf k}_T^2)^2 {\mathbf k}_T^2 } F_{\psi}({\mathbf k}_T^2), \eqno (A.3)
$$

\noindent
where $|\psi(0)|^2 = 0.0876\,{\rm GeV}^3$ is the $J/\psi$ wave function at the
origin, $m_\psi$ is the $J/\psi$ meson mass, ${\mathbf k}_T^2$ is the virtuality of 
the initial gluon, and function $F_{\psi}({\mathbf k}_T^2)$ is given by
$$
  F_{\psi}({\mathbf k}_T^2) = {\mathbf k}_T^2({\mathbf k}_T^6(m_\psi^2 - \hat t)(m_\psi^2 - \hat t - \hat u) - m_\psi^2 ({\hat t}^2 + \hat t \hat u + {\hat u}^2 - 
$$
$$
  m_\psi^2(\hat t + \hat u))^2 +  {\mathbf k}_T^4(3m_\psi^6 + \hat t\hat u(\hat t + \hat u) - 3m_\psi^4(2\hat t + \hat u) + m_\psi^2(3{\hat t}^2 + 2\hat t\hat u - {\hat u}^2)) + 
$$
$$
  {\mathbf k}_T^2(2m_\psi^8 + m_\psi^4\hat t(\hat t - \hat u) - {\hat t}^2 (\hat t + \hat u)^2 - 2m_\psi^6(2\hat t + \hat u) + m_\psi^2\hat t (2{\hat t}^2 + 5\hat t\hat u + 5{\hat u}^2))) + 
$$
$$
  2{\mathbf k}_T^2({\mathbf k}_T^4(m_\psi^2 - \hat t)(m_\psi^2 - \hat t - u) + {\mathbf k}_T^2(3m_\psi^6 + \hat t(\hat t + \hat u)^2 - m_\psi^4(5\hat t + 3\hat u) + 
$$
$$
  m_\psi^2({\hat t}^2 + \hat t\hat u - 2{\hat u}^2)) + m_\psi^2(2m_\psi^6 - m_\psi^4(3\hat t + 2\hat u) + 
$$
$$
  \hat t({\hat t}^2 + 2\hat t\hat u + 3{\hat u}^2)))( - |{\mathbf p}_{\psi T}||{\mathbf k}_T|\cos\phi_2) + 2m_\psi^2(m_\psi^2(-m_\psi^2 + \hat t + \hat u)^2 + 
$$
$$
  {\mathbf k}_T^2(m_\psi^4 + {\hat t}^2 + 2\hat t\hat u - {\hat u}^2 - 2m_\psi^2(\hat t + \hat u))){\mathbf p}_{\psi T}^2{\mathbf k}_T^2\cos^2\phi_2. \eqno (A.4)
$$

\noindent
Here ${\mathbf p}_{\psi T}$ is the $J/\psi$ transverse momentum, $\phi_2$ is the 
azimutal angle of the incoming virtual gluon having virtuality ${\mathbf k}_T^2$.
To study polarized $J/\psi$ production we introduce the four-vector of the 
longitudinal polarization $\epsilon_{\psi, L}^{\mu}$. In the frame where the $z$ axis
is oriented along the quarkonium momentum vector, 
$p_\psi^\mu = (E_\psi,0,0,|{\mathbf p}_\psi|)$, this polarization vector is
$\epsilon_{\psi, L}^\mu = (|{\mathbf p}_\psi|,0,0,E_\psi)/m_\psi$. The 
squared off-shell matrix element read
$$
  |\bar {\cal M}|_L^2(\gamma g^* \to J/\psi g) = {32 e_Q^2 (4\pi)^3 \alpha \alpha_s^2 |\psi(0)|^2 m_{\psi} \over 
  3 (\hat t - m_\psi^2)^2 (\hat u - m_\psi^2 - {\mathbf k}_T^2)^2 (\hat t + \hat u + {\mathbf k}_T^2)^2 {\mathbf k}_T^2 } F_{\psi, L}({\mathbf k}_T^2), \eqno (A.5)
$$

\noindent where function $F_{\psi, L}({\mathbf k}_T^2)$ is defined as
$$
  F_{\psi, L}({\mathbf k}_T^2) = -4{\Delta}_2{\Delta}_3{\mathbf k}_T^4 m_\psi^4 + 2 {\Delta}_3^2 {\mathbf k}_T^4 m_\psi^4 - 4 {\Delta}_2 {\Delta}_3 {\mathbf k}_T^2 m_\psi^6 + 
$$
$$
  8 {\Delta}_2 {\Delta}_3 {\mathbf k}_T^4 m_\psi^2\hat t - 4 {\Delta}_3^2 {\mathbf k}_T^4 m_\psi^2 \hat t + 8 {\Delta}_2 {\Delta}_3 {\mathbf k}_T^2 m_\psi^4 \hat t + 4 {\Delta}_3^2 {\mathbf k}_T^2 m_\psi^4 \hat t - 
$$
$$
  4 {\Delta}_2 {\Delta}_3 {\mathbf k}_T^4 {\hat t}^2 + 2 {\Delta}_3^2 {\mathbf k}_T^4 {\hat t}^2 - 4 {\Delta}_2 {\Delta}_3 {\mathbf k}_T^2 m_\psi^2 {\hat t}^2 - 8 {\Delta}_3^2 {\mathbf k}_T^2 m_\psi^2 {\hat t}^2 + 2 {\Delta}_3^2 m_\psi^4 {\hat t}^2 + 
$$
$$
  4 {\Delta}_3^2 {\mathbf k}_T^2 {\hat t}^3 - 4 {\Delta}_3^2 m_\psi^2 {\hat t}^3 + 2 {\Delta}_3^2 {\hat t}^4 + 4 {\Delta}_2^2 {\mathbf k}_T^4 m_\psi^2 \hat u - 
$$
$$
  4 {\Delta}_2 {\Delta}_3 {\mathbf k}_T^4 m_\psi^2 \hat u + 2 {\Delta}_3^2 {\mathbf k}_T^4 m_\psi^2 \hat u + 4 {\Delta}_2^2 {\mathbf k}_T^2 m_\psi^4 \hat u + 2 {\Delta}_3^2 {\mathbf k}_T^2 m_\psi^4 \hat u - 
$$
$$
  4 {\Delta}_2^2 {\mathbf k}_T^4 \hat t \hat u + 4 {\Delta}_2 {\Delta}_3 {\mathbf k}_T^4 \hat t \hat u - 2 {\Delta}_3^2 {\mathbf k}_T^4 \hat t \hat u - 4 {\Delta}_2^2 {\mathbf k}_T^2 m_\psi^2\hat t\hat u - 4 {\Delta}_2 {\Delta}_3 {\mathbf k}_T^2 m_\psi^2\hat t\hat u - 
$$
$$
  4 {\Delta}_3^2 {\mathbf k}_T^2 m_\psi^2\hat t\hat u + 2 {\Delta}_3^2 m_\psi^4\hat t\hat u + 4 {\Delta}_2 {\Delta}_3 {\mathbf k}_T^2 {\hat t}^2\hat u + 2 {\Delta}_3^2 {\mathbf k}_T^2 {\hat t}^2\hat u - 6 {\Delta}_3^2 m_\psi^2 {\hat t}^2\hat u + 
$$
$$
  4 {\Delta}_3^2 {\hat t}^3\hat u + {\mathbf k}_T^6 {\hat u}^2 - 4 {\Delta}_2^2 {\mathbf k}_T^2 m_\psi^2 {\hat u}^2 + 4 {\Delta}_2 {\Delta}_3 {\mathbf k}_T^2 m_\psi^2 {\hat u}^2 - 2 {\Delta}_3^2 {\mathbf k}_T^2 m_\psi^2 {\hat u}^2 + {\mathbf k}_T^2 m_\psi^4 {\hat u}^2 + 
$$
$$
  4 {\Delta}_2 {\Delta}_3 {\mathbf k}_T^2\hat t {\hat u}^2 - 2 {\Delta}_3^2 {\mathbf k}_T^2\hat t {\hat u}^2 - 2 {\mathbf k}_T^4\hat t {\hat u}^2 - 2 {\Delta}_3^2 m_\psi^2\hat t {\hat u}^2 - 2 {\mathbf k}_T^2 m_\psi^2\hat t {\hat u}^2 + 
$$
$$
  2 {\Delta}_3^2 {\hat t}^2 {\hat u}^2 + 2 {\mathbf k}_T^2 {\hat t}^2 {\hat u}^2 - 2 {\mathbf k}_T^2 m_\psi^2 {\hat u}^3 + 2 {\mathbf k}_T^2\hat t {\hat u}^3 + {\mathbf k}_T^2 {\hat u}^4 + 8 {\Delta}_1^2 {\mathbf k}_T^4 (-m_\psi^2 + \hat t + \hat u)^2 + 
$$
$$
  4 {\Delta}_1 {\mathbf k}_T^2 ({\Delta}_2 ({\mathbf k}_T^4 (2 m_\psi^2 - 2\hat t - \hat u) + m_\psi^2 (m_\psi^2 - 2\hat t - \hat u)\hat u + {\mathbf k}_T^2 (2 m_\psi^4 + {\hat u}^2 - 
$$
$$
  2 m_\psi^2 (\hat t + \hat u))) + {\Delta}_3 (-m_\psi^6 + {\mathbf k}_T^4 (-m_\psi^2 + \hat t) + 2\hat t (\hat t + \hat u)^2 + 
$$
$$
  m_\psi^4 (3\hat t + \hat u) - m_\psi^2\hat t (4\hat t + 3\hat u) + {\mathbf k}_T^2 (-(m_\psi^2 (2\hat t + \hat u)) + \hat t (2\hat t + 3\hat u)))) + 
$$
$$
  4 (-({\Delta}_2 {\Delta}_3 ({\mathbf k}_T^2 + m_\psi^2) (m_\psi^2 - \hat t) (m_\psi^2 - \hat t - \hat u)) + {\mathbf k}_T^2 ({\mathbf k}_T^2 - \hat t) {\hat u}^2 + 
$$
$$
  4 {\Delta}_1^2 {\mathbf k}_T^2 (-m_\psi^2 + \hat t + \hat u)^2 + {\Delta}_1 (2 {\Delta}_2 {\mathbf k}_T^2 (2 m_\psi^4 + {\mathbf k}_T^2 (2 m_\psi^2 - 2\hat t - \hat u) - 
$$
$$
  2 m_\psi^2 (\hat t + \hat u) + \hat u (\hat t + \hat u)) - {\Delta}_3 (m_\psi^6 + {\mathbf k}_T^4 (m_\psi^2 - \hat t + \hat u) - 
$$
$$
  2\hat t (\hat t + \hat u)^2 - m_\psi^4 (3\hat t + 2\hat u) + m_\psi^2 (4 {\hat t}^2 + 5\hat t\hat u + {\hat u}^2) + 
$$
$$
  {\mathbf k}_T^2 (-2 {\hat t}^2 - t\hat u + {\hat u}^2 + m_\psi^2 (2\hat t + \hat u))))) ( - |{\mathbf p}_{\psi T}||{\mathbf k}_T|\cos\phi_2) + 
$$
$$
  4 (2 {\Delta}_1 {\Delta}_2 ({\mathbf k}_T^2 + m_\psi^2) (m_\psi^2 - \hat t - \hat u) + 
$$
$$
  {\mathbf k}_T^2 {\hat u}^2 + 2 {\Delta}_1^2 (-m_\psi^2 + \hat t + \hat u)^2) {\mathbf p}_{\psi T}^2{\mathbf k}_T^2\cos^2\phi_2, \eqno (A.6)
$$

\noindent
and the following notation has been used:
$$
  \Delta_1 = (\alpha_1 + \alpha_2){\sqrt s\over 2m_\psi}\left(\sqrt{{\mathbf p}_{\psi T}^2 + s(\alpha_1 - \beta_1)^2/4} - {s(\alpha_1^2 - \beta_1^2)\over 4\sqrt{{\mathbf p}_{\psi T}^2 + s(\alpha_1 - \beta_1)^2/4}}\right),
$$
$$
  \displaystyle \Delta_2 = (\beta_1 + \beta_2){\sqrt s\over 2m_\psi}\left(\sqrt{{\mathbf p}_{\psi T}^2 + s(\alpha_1 - \beta_1)^2/4} + {s(\alpha_1^2 - \beta_1^2)\over 4\sqrt{{\mathbf p}_{\psi T}^2 + s(\alpha_1 - \beta_1)^2/4}}\right) + \Delta_3,
$$
$$
  \Delta_3 = - {\sqrt{s} (\alpha_1 + \beta_1) \over 2m_\psi \sqrt{{\mathbf p}_{\psi T}^2 + s(\alpha_1 - \beta_1)^2/4}} |{\mathbf p}_{\psi T}| |{\mathbf k}_{T}| \cos\phi_2, 
$$
$$
  \alpha_1 = \sqrt{ m_\psi^2 + {\mathbf p}_{\psi T}^2\over s} \exp (y_\psi), \quad \beta_1 = \sqrt{ m_\psi^2 + {\mathbf p}_{\psi T}^2\over s} \exp ( - y_\psi),
$$
$$
  \alpha_2 = {|{\mathbf p}_{g T}| \over \sqrt s} \exp (y_g), \quad \beta_2 = {|{\mathbf p}_{g T}| \over \sqrt s} \exp (-y_g). \eqno(A.7)
$$

\begin{figure}
\begin{center}
\epsfig{figure=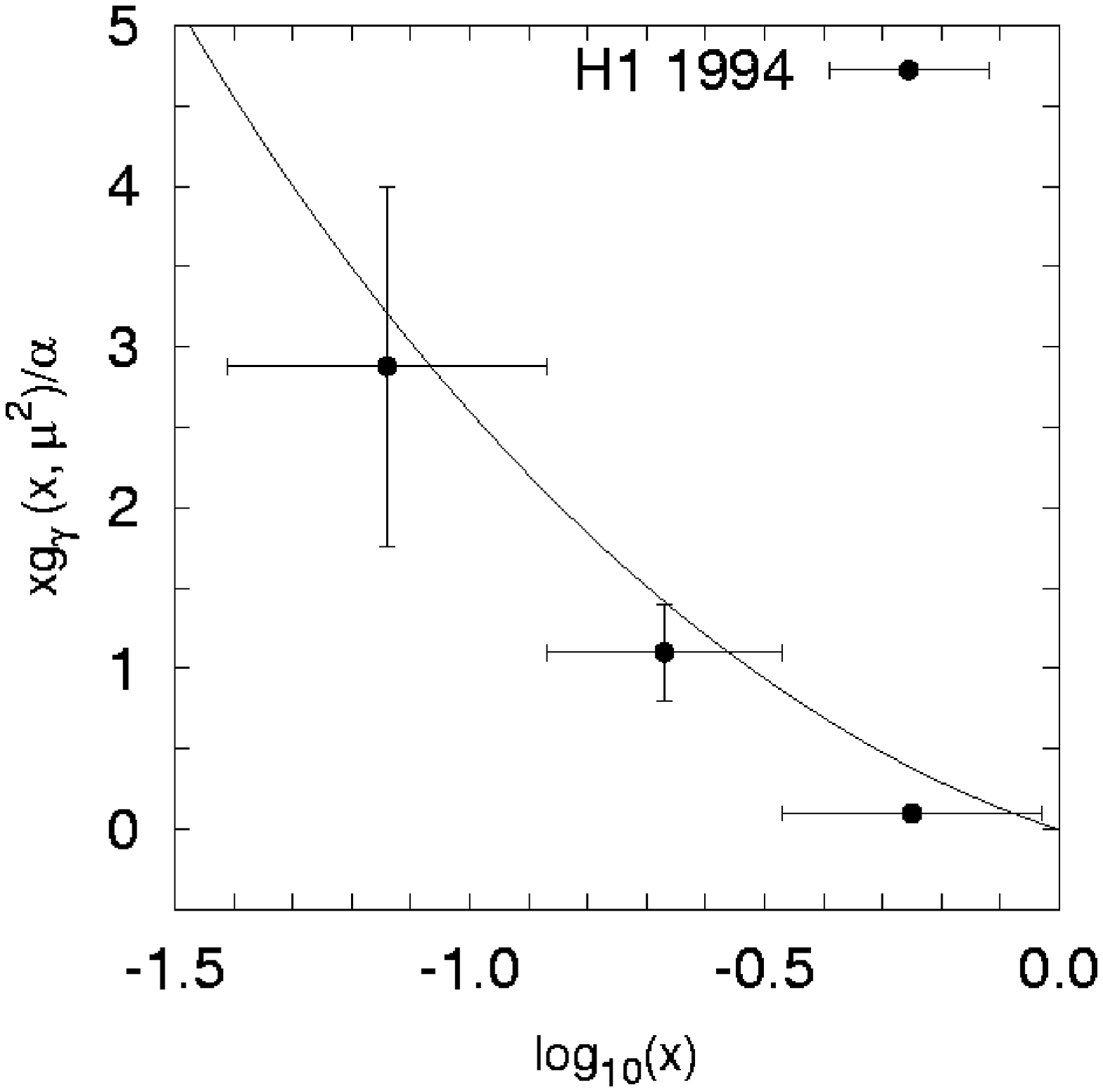}
\end{center}
\caption{The effective gluon distribution $xg_\gamma(x,\mu^2)$ in the photon as a 
  function of ${\rm log}_{10} x$ at $\mu^2 = 38\,{\rm GeV^2}$. The experimental data~[38] taken by H1 
  collaboration from hard dijets production at HERA.}
\label{fig1}
\end{figure}

\newpage

\begin{figure}
\begin{center}
\epsfig{figure=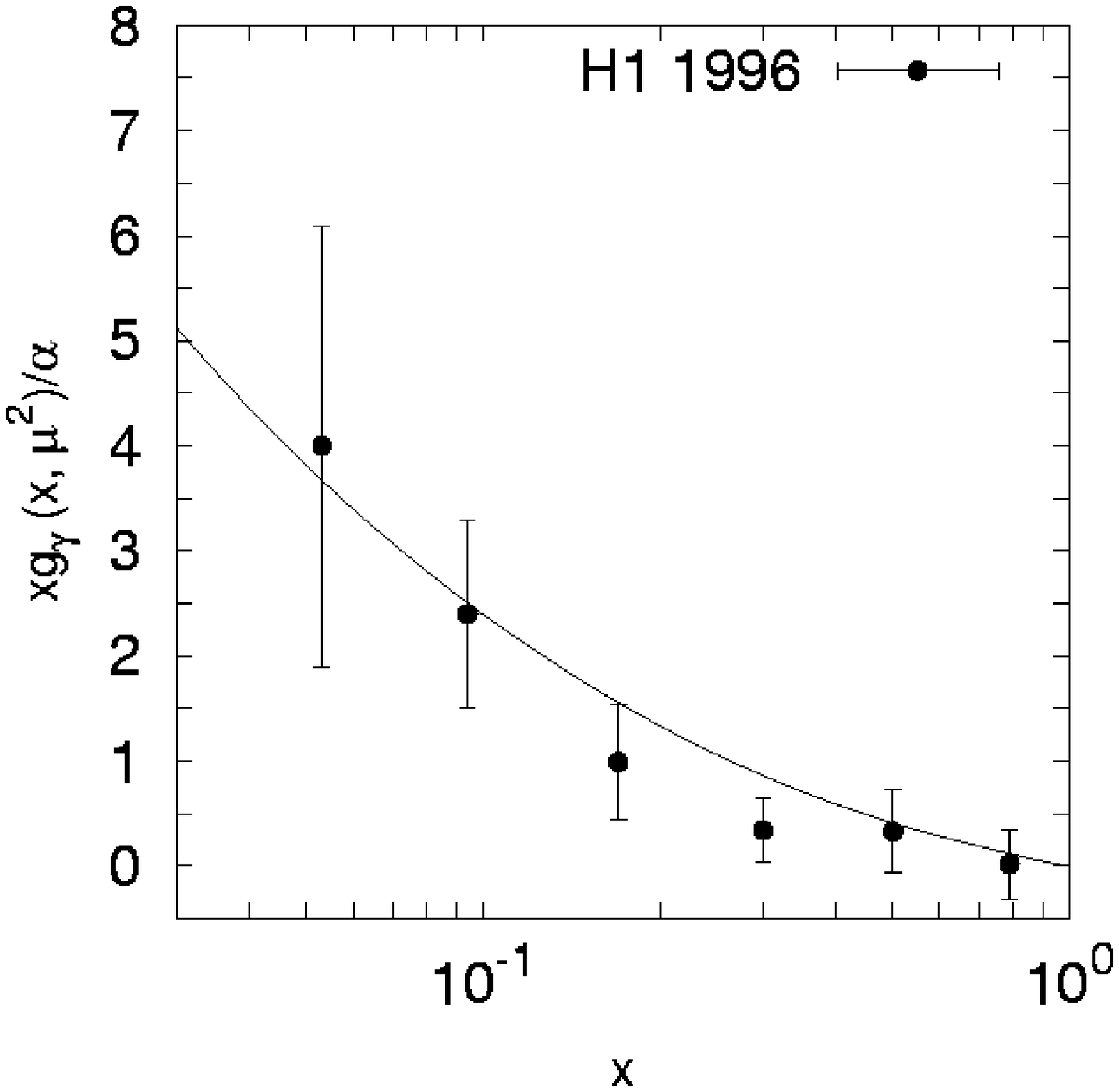}
\end{center}
\caption{The effective gluon distribution $xg_\gamma(x,\mu^2)$ in the photon as a 
  function of $x$ at $\mu^2 = 74\,{\rm GeV^2}$. The experimental data~[39] taken by H1 
  collaboration from hard dijets production at HERA.}
\label{fig2}
\end{figure}

\newpage

\begin{figure}
\begin{center}
\epsfig{figure=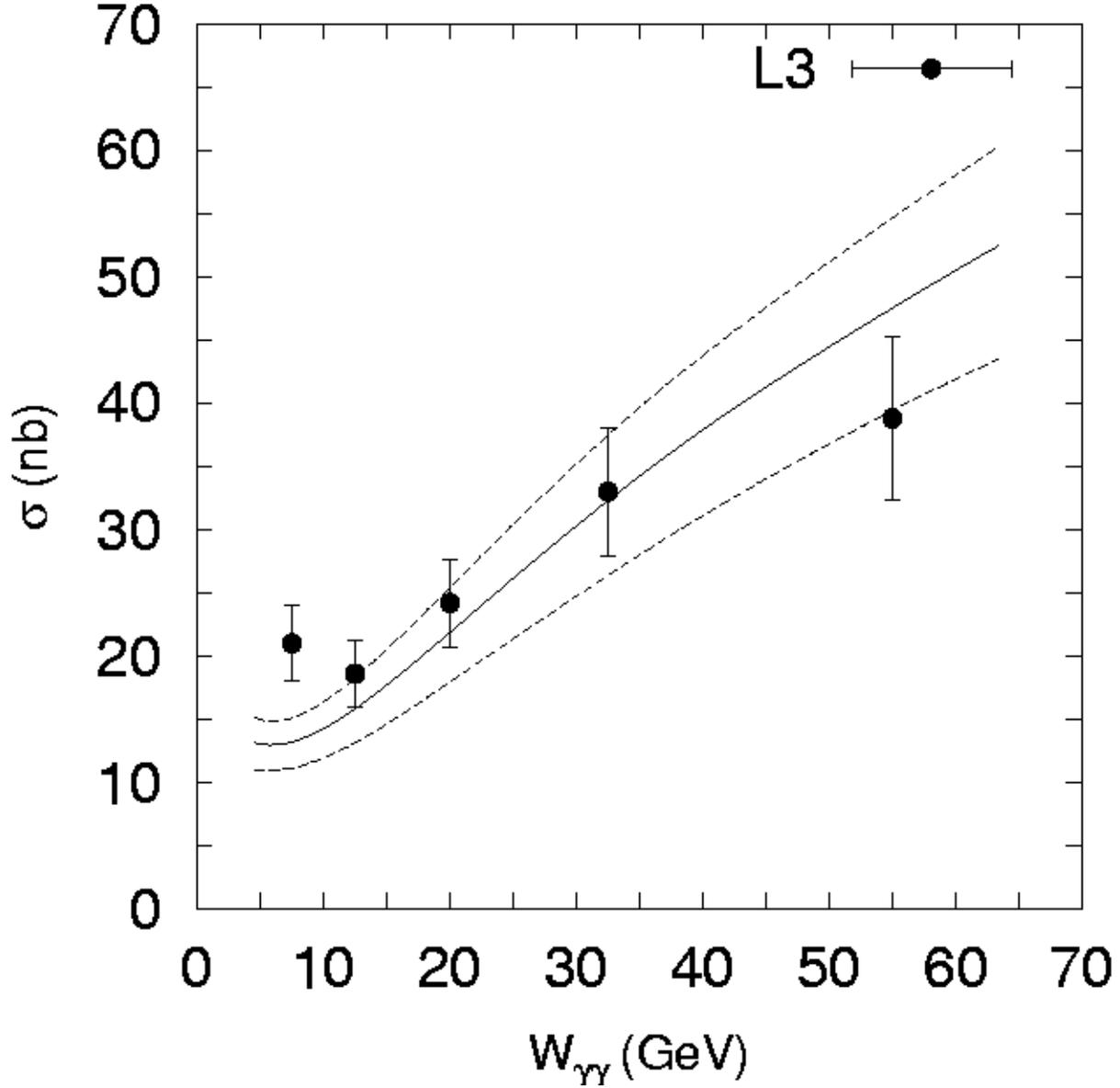}
\end{center}
\caption{The open charm total cross section $\sigma(\gamma\gamma \to c\bar c + X)$ as a function of
  $W_{\gamma \gamma}$ at $\sqrt s = 189 - 202$ GeV. The solid line corresponds to
  the charm mass $m_c = 1.4$ GeV, upper and lower dashed lines 
  correspond to the $m_c = 1.3$ GeV and $m_c = 1.5$ GeV respectively. The experimental 
  data~[2] are taken by the L3 collaboration.}
\label{fig3}
\end{figure}

\newpage

\begin{figure}
\begin{center}
\epsfig{figure=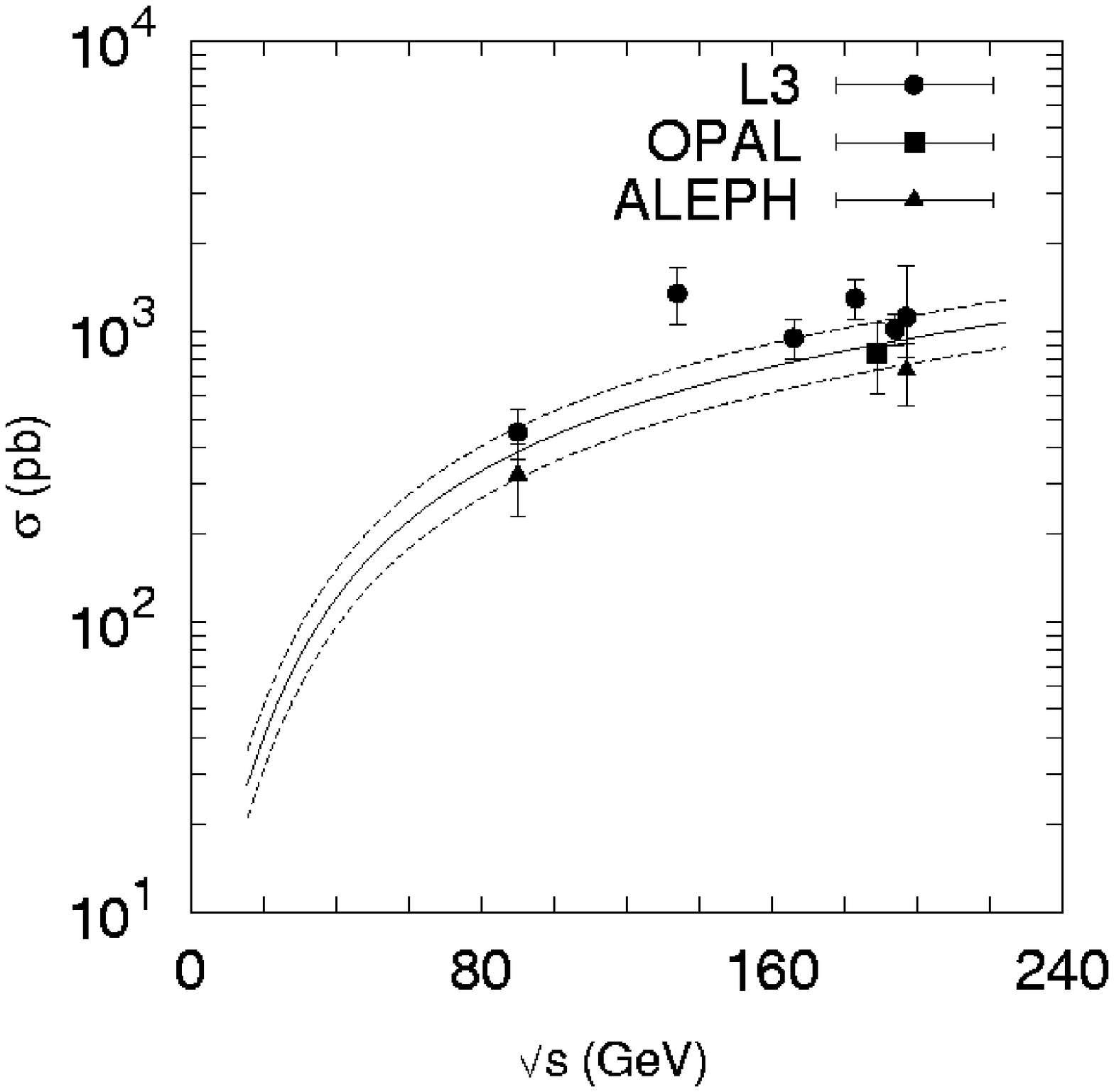}
\end{center}
\caption{The open charm total cross section $\sigma(e^+ e^- \to e^+ e^- c\bar c + X)$ as a function of
  the $e^+ e^-$ centre-of-mass energy $\sqrt s$. All curves are the same as Figure 3. 
  The experimental data are from L3~[1, 3], OPAL~[4] and ALEPH~[5] collaborations.}
\label{fig4}
\end{figure}

\newpage

\begin{figure}
\begin{center}
\epsfig{figure=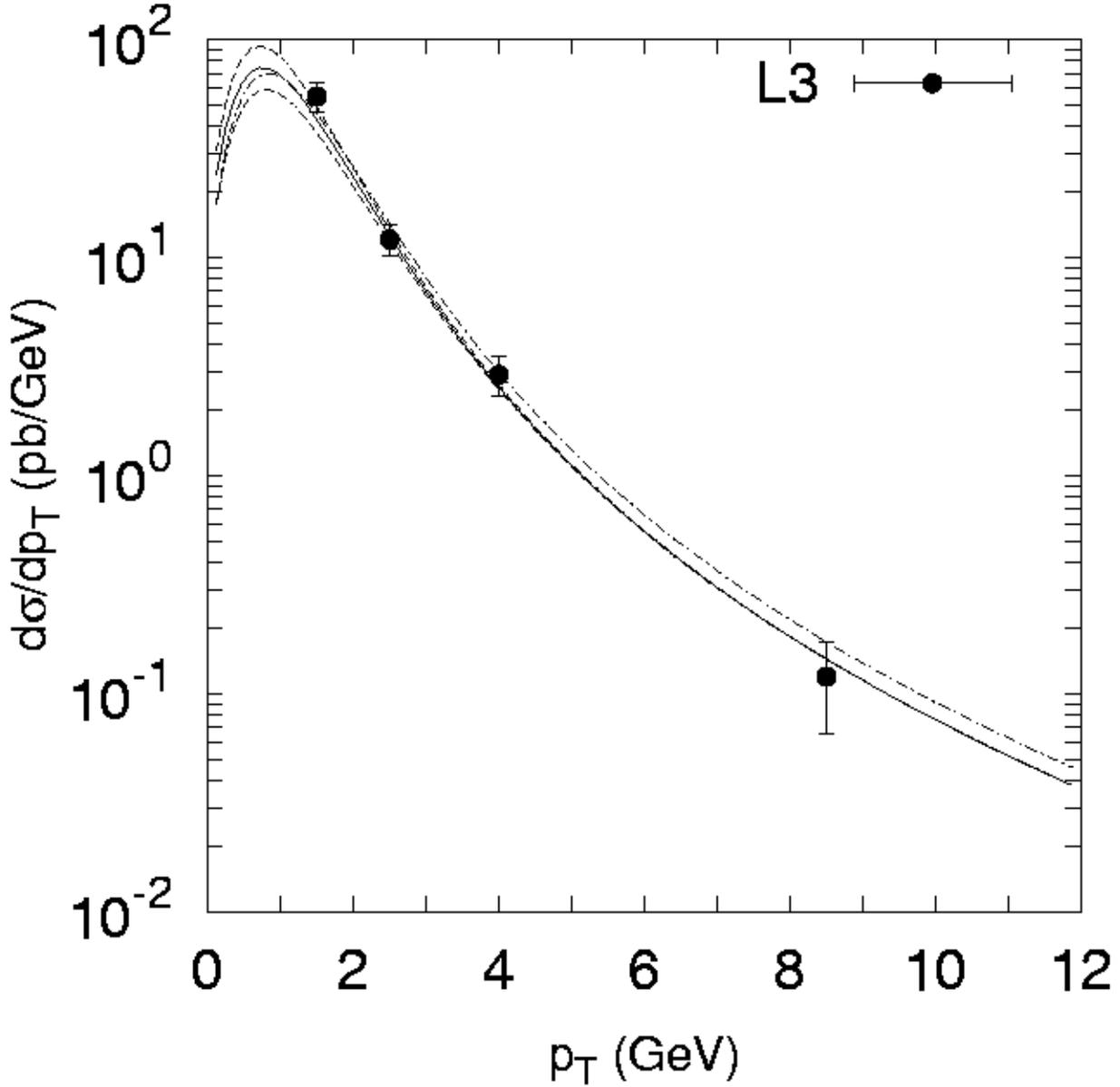}
\end{center}
\caption{The differential cross section $d\sigma/dp_T$ for inclusive $D^*$ production at
  $|\eta| < 1.4$. The solid and both dashed curves here are the same as in Figure 3
  (calculated with the default value $\epsilon_c = 0.06$),
  dash-dotted curve represents results obtained using $\epsilon_c = 0.031$ and $m_c = 1.4$ GeV.
  The experimental data are from L3~[3].}
\label{fig5}
\end{figure}

\newpage

\begin{figure}
\begin{center}
\epsfig{figure=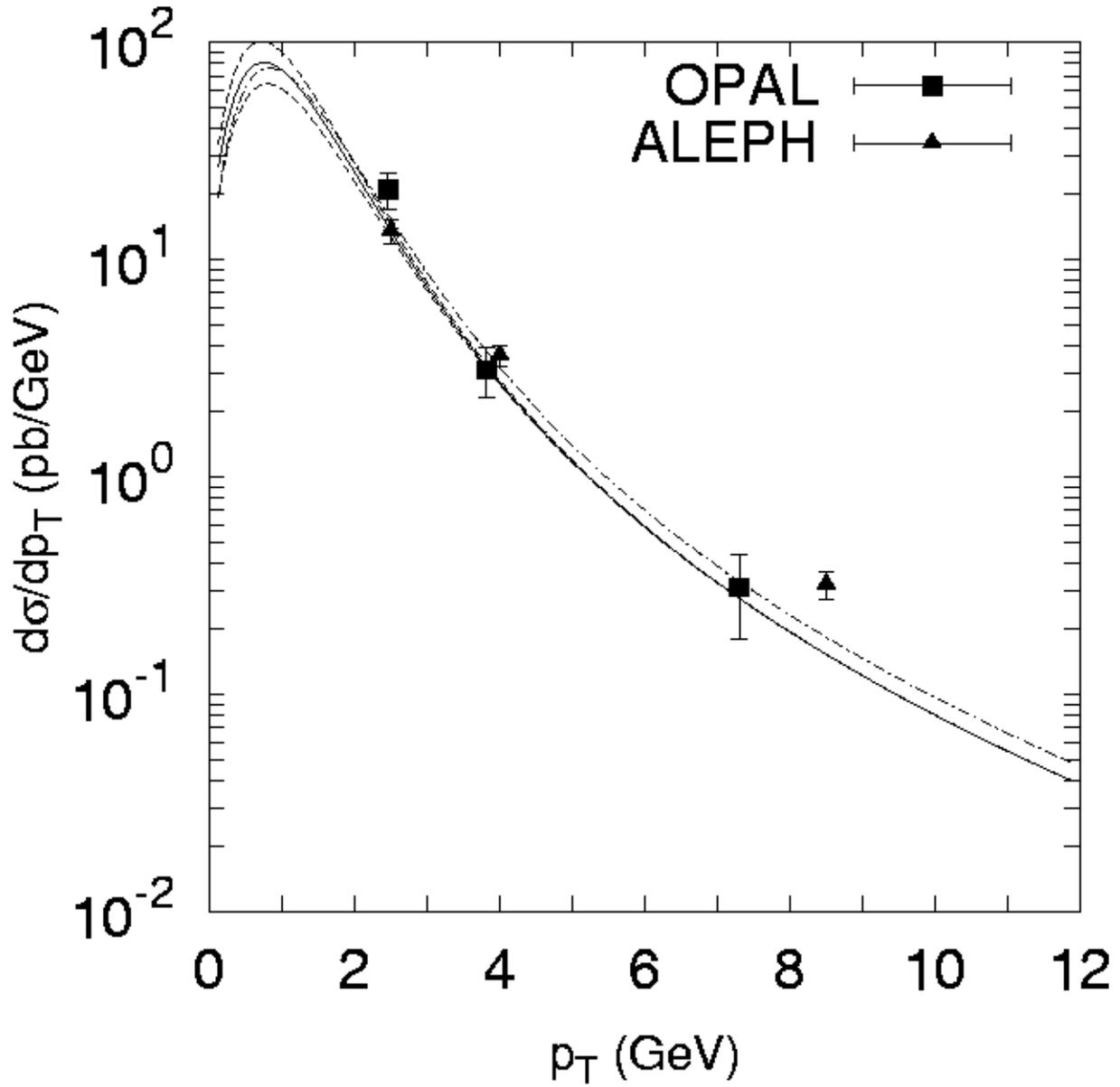}
\end{center}
\caption{The differential cross section $d\sigma/dp_T$ for inclusive $D^*$ production
  at $|\eta| < 1.5$. All curves here are the same as in Figure 5. The experimental data are from 
  OPAL~[4] and ALEPH~[5].}
\label{fig6}
\end{figure}

\newpage

\begin{figure}
\begin{center}
\epsfig{figure=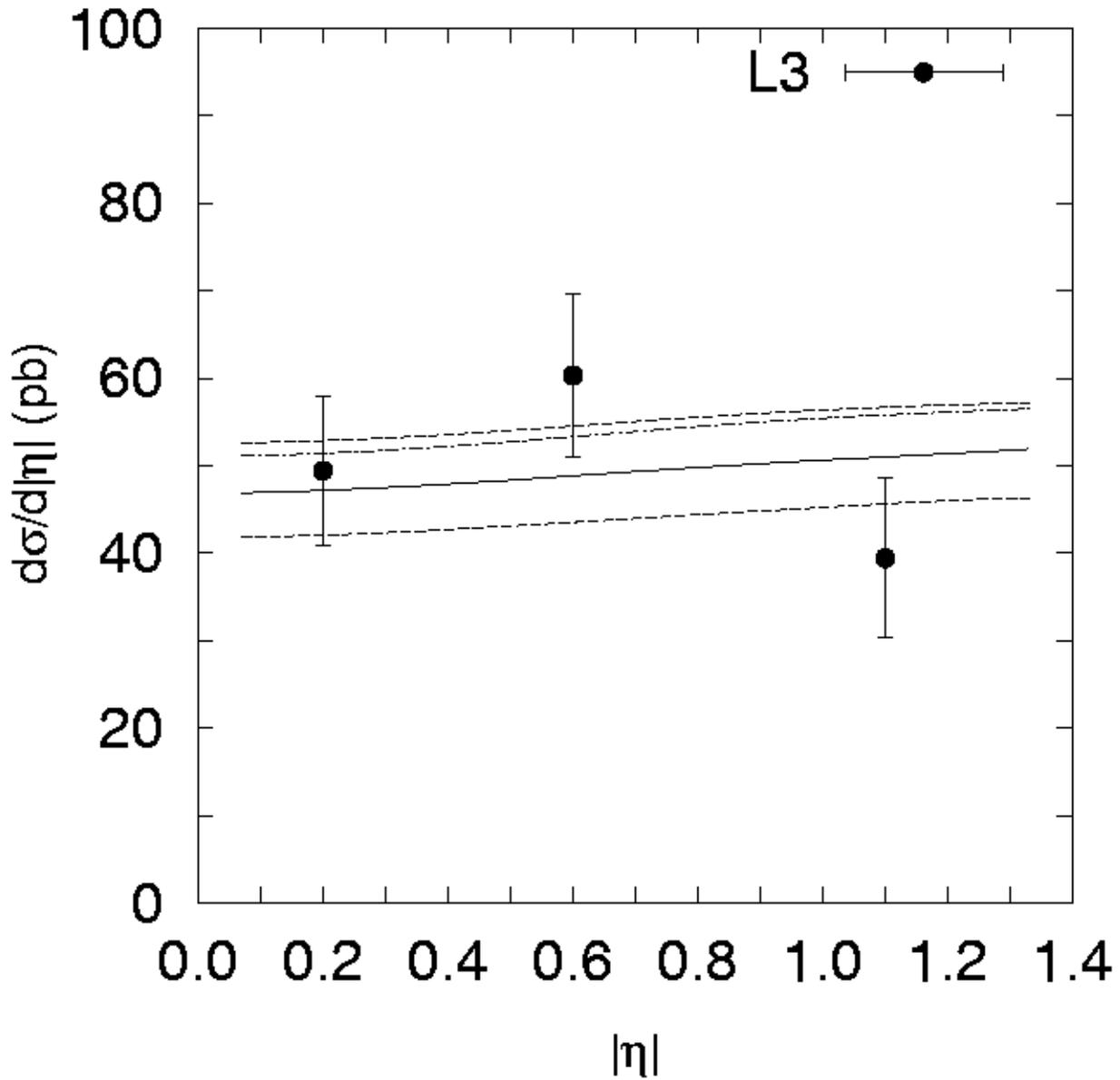}
\end{center}
\caption{The differential $D^*$ cross section $d\sigma/d|\eta|$ for the process 
  $e^+ e^- \to e^+ e^- D^* + X$ in the $1 < p_T < 12$ GeV range. 
  All curves here are the same as in Figure 5. The experimental data are from L3~[3].}
\label{fig7}
\end{figure}

\newpage

\begin{figure}
\begin{center}
\epsfig{figure=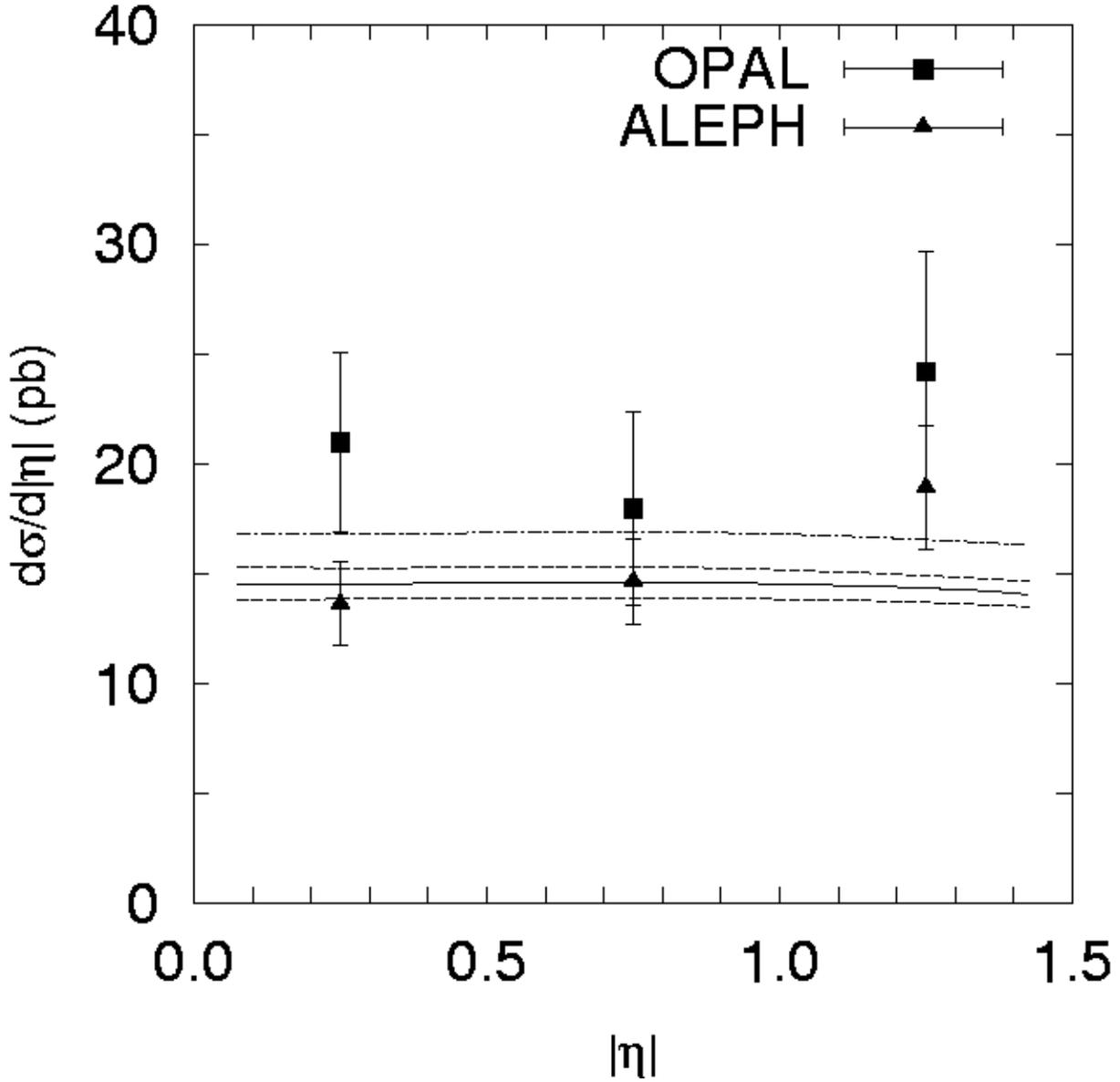}
\end{center}
\caption{The differential $D^*$ cross section $d\sigma/d|\eta|$ for the process 
  $e^+ e^- \to e^+ e^- D^* + X$ in the $2 < p_T < 12$ GeV range. 
  All curves here are the same as in Figure 5. The experimental data are from 
  OPAL~[4] and ALEPH~[5] collaborations.}
\label{fig8}
\end{figure}

\newpage

\begin{figure}
\begin{center}
\epsfig{figure=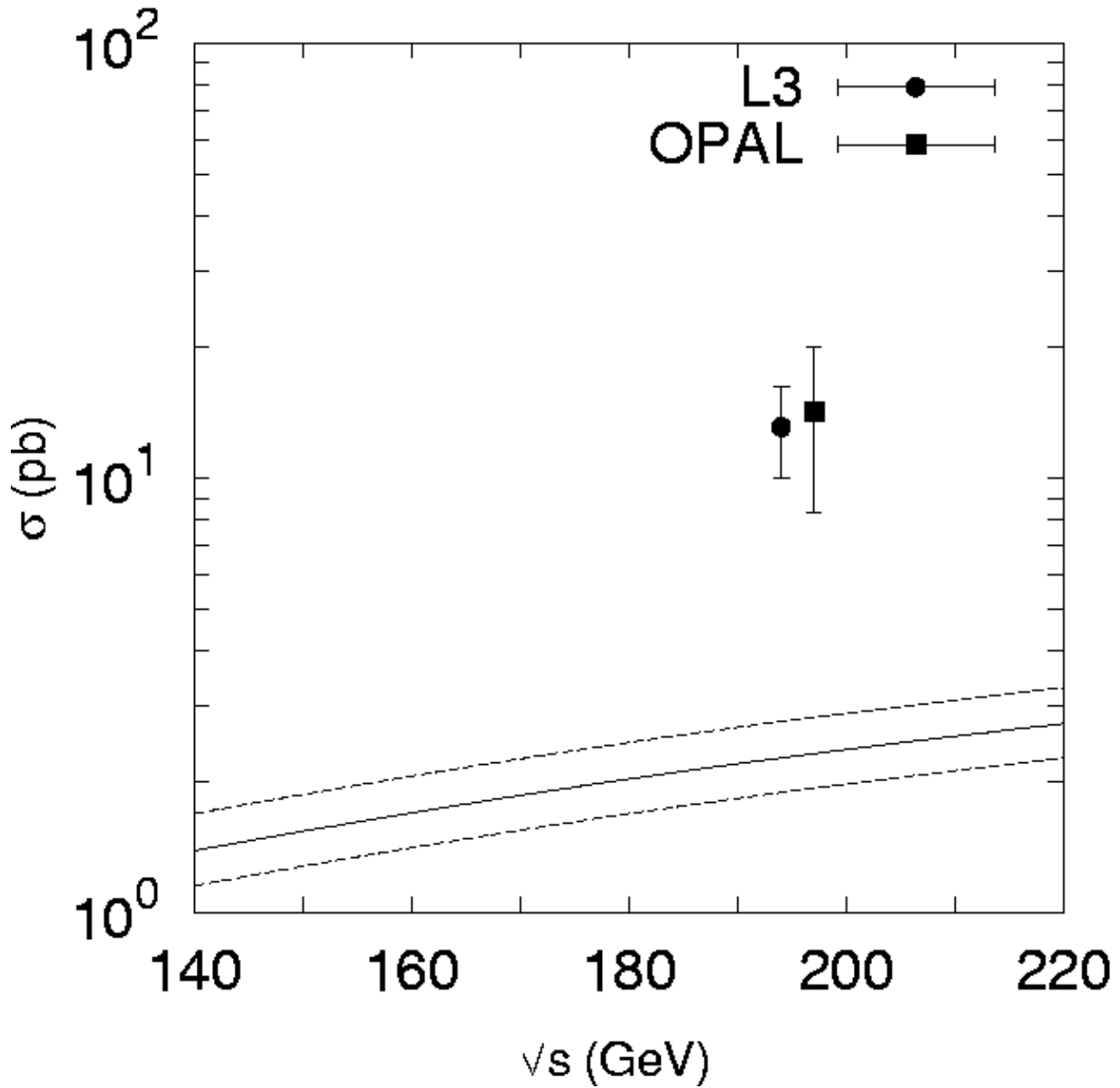}
\end{center}
\caption{The bottom total cross section $\sigma(e^+ e^- \to e^+ e^- b\bar b + X)$ as a function of
  the $e^+ e^-$ centre-of-mass energy $\sqrt s$. 
  The solid line corresponds to the bottom mass $m_b = 4.75$ GeV, upper and lower dashed lines 
  correspond to the $m_b = 4.5$ GeV and $m_b = 5.0$ GeV respectively.
  The experimental data are from L3~[1] and OPAL~[4].}
\label{fig9}
\end{figure}

\newpage

\begin{figure}
\begin{center}
\epsfig{figure=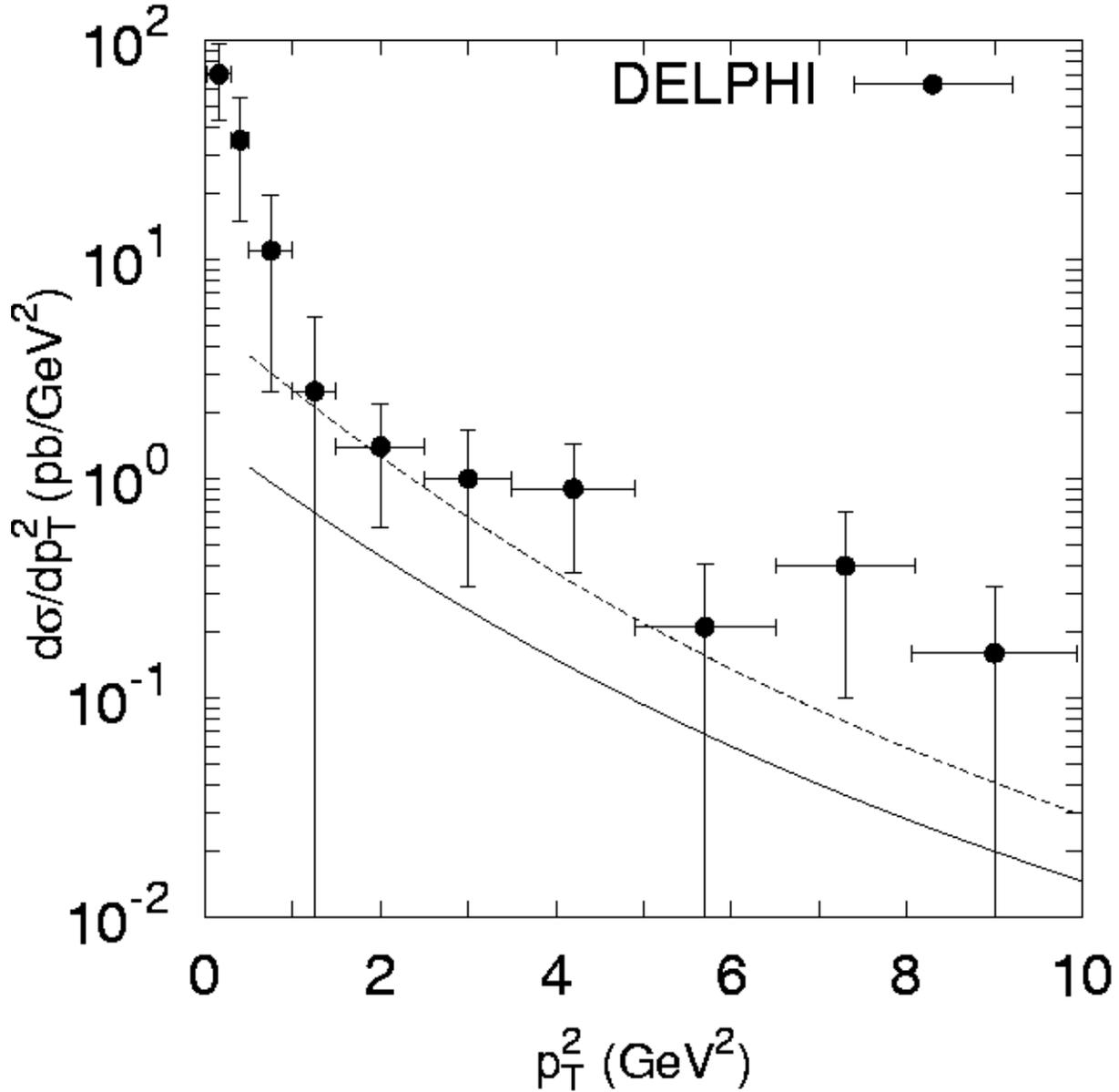}
\end{center}
\caption{The cross section $d\sigma/dp_{\psi T}^2$ of $e^+ e^- \to e^+ e^- J/\psi + X$ measured 
  by DELPHI collaboration~[35] at $2 < y_\psi < 2$ as a function of $p_{\psi T}^2$ is compared
  with the $k_T$-factorization calculations in the CS model.
  The solid line corresponds to the $m_c = 1.55$ GeV and $\Lambda_{\rm QCD} = 200$ MeV,
  dashed line corresponds to the $m_c = 1.4$ GeV and $\Lambda_{\rm QCD} = 340$ MeV.}
\label{fig10}
\end{figure}

\newpage

\begin{figure}
\begin{center}
\epsfig{figure=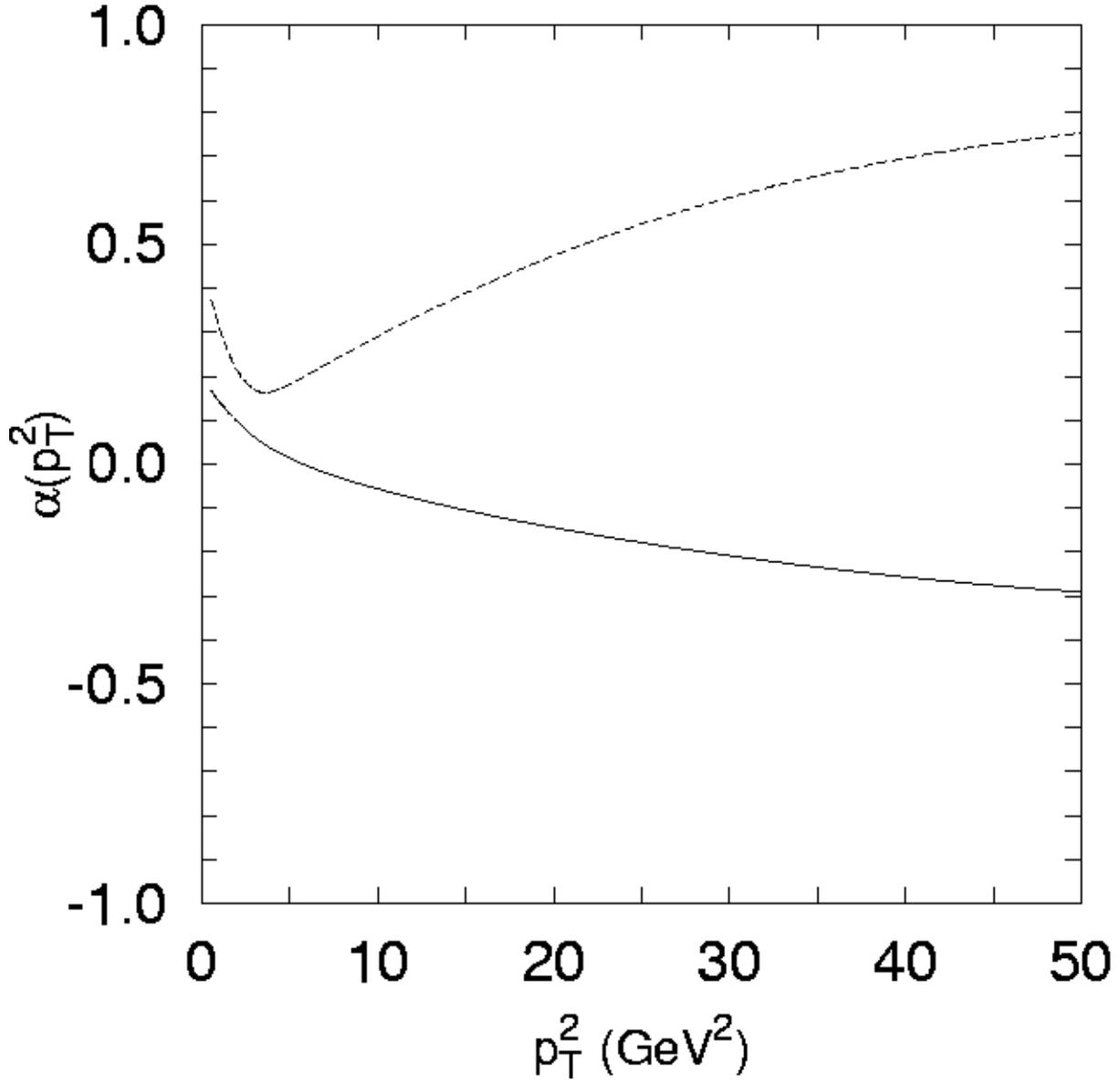}
\end{center}
\caption{The $p_{\psi T}^2$ dependence of the spin alignement parameter $\alpha$ for 
  the inclusive $J/\psi$ production. The solid line corresponds to the $k_T$-factorization
  predictions, dashed line corresponds to the collinear leading order pQCD calculations with 
  the GRV (LO) gluon density in a photon.}
\label{fig11}
\end{figure}

\end{document}